\documentclass[journal=acsodf,manuscript=article,layout=twocolumn]{achemso}
\usepackage[version=4]{mhchem}
\usepackage{graphicx}
\usepackage{xcolor}
\usepackage{multirow}
\usepackage{amsmath}
\usepackage{array}
\usepackage{stfloats}
\usepackage{placeins}

\title{Comprehensive First-Principles Investigation of the Structural, Mechanical, Electronic, and Optical Properties of Homoelemental Phase T-GaN Monolayer}

\author{Djardiel S. Gomes}
\affiliation{Princesa Isabel Campus, Federal Institute of Education, Science and Technology of Para'iba (IFPB), Princesa Isabel, PB, 58755-000, Brazil.}
\alsoaffiliation{Materials Science Postgraduate Program, College UnB Planaltina, University of Bras\'{i}lia, Bras\'{i}lia, DF, 73380-900, Brazil.}

\author{Isaac M. F\'{e}lix}
\affiliation{Center for Agri-food Science and Technology, Federal University of Campina Grande, Pombal, PB, 58840-000, Brazil.}
\alsoaffiliation{Department of Physics, Federal University of Campina Grande, Campina Grande, PB, 58429-900, Brazil.}

\author{Jorge O. A. L. Torres}
\affiliation{Department of Electrical Engineering, College of Technology, University of Bras\'{i}lia, Bras\'{i}lia, DF, 70910-900, Brazil.}

\author{\\Fabio L. L. de Mendonça}
\affiliation{Department of Electrical Engineering, College of Technology, University of Bras\'{i}lia, Bras\'{i}lia, DF, 70910-900, Brazil.}

\author{Sergio Azevedo}
\affiliation{Department of Physics, Federal University of Para\'{i}ba, Jo\~{a}o Pessoa, PB, 58051-900, Brazil.}

\author{Marcelo L. Pereira Junior}
\email{marcelo.lopes@unb.br}
\affiliation{Department of Electrical Engineering, College of Technology, University of Bras\'{i}lia, Bras\'{i}lia, DF, 70910-900, Brazil.}
\alsoaffiliation{Materials Science Postgraduate Program, College UnB Planaltina, University of Bras\'{i}lia, Bras\'{i}lia, DF, 73380-900, Brazil.}

\begin{document}
	
	\begin{abstract}
		The exploration of non-hexagonal two-dimensional topologies has opened new possibilities for tailoring the properties of group III-V monolayers beyond those accessible through conventional honeycomb phases. In this context, we have investigated the structural, mechanical, electronic, and optical properties of T-GaN, a two-dimensional tetragonal gallium nitride monolayer composed of alternating four- and eight-membered rings featuring coexisting homoelemental (Ga-Ga, N-N) and heteropolar (Ga-N) bonds, using density functional theory (DFT) within the generalized gradient approximation (GGA/PBE) and the hybrid HSE06 functional. The dynamical stability of T-GaN was confirmed by phonon dispersion calculations, which revealed the absence of imaginary frequencies throughout the Brillouin zone, and was further supported by \textit{ab initio} molecular dynamics (AIMD) simulations. The mechanical characterization reveals a pronounced in-plane anisotropy, with critical strains of approximately 16.5\% and 7.0\% along the $x$- and $y$-directions, respectively. The electronic band structure analysis indicates that T-GaN is a nonmagnetic semiconductor with an indirect band gap of 0.35~eV (PBE) and 1.15~eV (HSE06), with the valence band maximum dominated by nitrogen 2\textit{p} orbitals and the conduction band minimum governed by gallium 4\textit{s} and 4\textit{p} states. The optical response was evaluated along three crystallographic directions, exhibiting considerable anisotropy in the absorption coefficient, refractive index, and reflectivity. These findings provide new insights into the physical properties of tetragonal group III-V monolayers and suggest that T-GaN may serve as a promising candidate for anisotropic nanoelectronic and optoelectronic applications.
	\end{abstract}
	
	%%%%%%%%%%%%%%%%%%%%%%%%%%%%%%%%%%%%%%%%%%%%%%%%%%%%%%%%%%%%%%%%%%%%%%%%%%%%%
	\section{Introduction}
	
	Two-dimensional (2D) materials have attracted considerable attention over the past two decades owing to their combination of structural, mechanical, electronic, and optical properties that differ substantially from those of their bulk counterparts \cite{Butler2013, Bhimanapati2015, Novoselov2016}. The isolation of graphene in 2004 \cite{Novoselov2004} marked a turning point in materials science and stimulated intensive research into a wide variety of 2D systems, including transition metal dichalcogenides \cite{Manzeli2017}, hexagonal boron nitride \cite{Cassabois2016}, and group III-V monolayers \cite{Zhuang2013}. Among these, gallium nitride (GaN) has emerged as a material of particular interest due to the technological relevance of its bulk wurtzite phase in optoelectronic devices such as light-emitting diodes, laser diodes, and ultraviolet photodetectors \cite{Morkoc1994, Nakamura1994}.
	
	The successful synthesis of 2D GaN was first achieved by Al Balushi \textit{et al.} \cite{AlBalushi2016} through migration-enhanced encapsulated growth on epitaxial graphene, and subsequently by Chen \textit{et al.} \cite{Chen2018} via surface-confined nitridation on liquid metals. These experimental achievements confirmed that atomically thin GaN is experimentally accessible and renewed interest in exploring the fundamental properties of GaN monolayers from a theoretical perspective \cite{Sanders2017, Sahu2023}. The hexagonal phase of monolayer GaN (g-GaN), which adopts a planar honeycomb-like structure with \textit{sp}$^{2}$ hybridization, has been extensively studied and is known to be an indirect band gap semiconductor with a gap of approximately 2.27~eV at the LDA/GGA level and 3.4--4.4~eV when corrected with hybrid functionals or GW methods \cite{Sahin2009, gGaN_bandgap_2, gGaN_bandgap_3}.
	
	Beyond hexagonal phases, the exploration of non-hexagonal 2D topologies has opened new avenues for tailoring material properties through structural diversity. In 2012, Liu \textit{et al.} \cite{Liu2012} proposed T-graphene, a carbon allotrope characterized by a periodic arrangement of four-membered and eight-membered rings with tetragonal symmetry, which was shown to host Dirac-like fermions with high Fermi velocity. This work inspired the theoretical prediction of a family of tetragonal 2D materials, including T-BN \cite{Wei2022}, T-SiC \cite{Yang2015_TSiC}, and T-AlN \cite{Luo2020}, all of which exhibited distinctive mechanical and electronic properties arising from their reduced symmetry compared to hexagonal phases.
	
	In the case of gallium nitride, several tetragonal phases have been proposed under the common designation of T-GaN, although they differ significantly in topology and bonding character. Yong \textit{et al.} \cite{Yong2017} investigated a tetragonal GaN monolayer with a square lattice in which each atom is four-fold coordinated to atoms of the opposite species, forming exclusively heteropolar Ga-N bonds within a network of four-membered rings, and reported a PBE band gap of approximately 1.88~eV along with promising gas sensing properties toward NO and NO$_2$ molecules. Zhang \textit{et al.} \cite{Zhang2017} characterized a distinct tetragonal GaN monolayer composed of alternating four- and eight-membered rings, also with exclusively Ga-N bonds, yielding PBE and HSE06 band gaps of approximately 1.89 and 3.12~eV, respectively. Camacho-Mojica and L\'{o}pez-Ur\'{i}as \cite{CamachoMojica2015} examined GaN Haeckelite nanostructures within the 4-8 ring topology, reporting an indirect band gap of approximately 1.6~eV at the PBE level for the monolayer configuration. Notably, tetragonal GaN configurations in which homoelemental Ga-Ga and N-N bonds coexist with heteropolar Ga-N bonds within the same 4-8 ring motif have received comparatively less attention, despite the fact that such mixed bonding environments may give rise to distinct mechanical and electronic properties. Despite these contributions, a comprehensive characterization of the structural stability, mechanical response, electronic structure, and optical properties of a T-GaN monolayer with mixed-bonding character is lacking in the literature. In particular, no previous study has simultaneously addressed these properties using both semilocal and hybrid functionals, nor explored the mechanical anisotropy and direction-dependent optical response of this monolayer.
	
	In this work, we have performed a comprehensive first-principles investigation of a T-GaN monolayer with mixed-bonding character within the 4-8 ring topology, using density functional theory (DFT) in both the GGA/PBE and HSE06 formalisms. The dynamical and thermal stability were assessed through phonon dispersion calculations and \textit{ab initio} molecular dynamics simulations (AIMD). The mechanical response was evaluated under uniaxial and biaxial strain. The electronic properties were characterized via band structure, density of states, projected density of states, spin-polarized calculations, and electron localization function analysis, and the optical properties were computed along three crystallographic directions. The results reported in this work provide a thorough physical description of T-GaN and offer insights into its potential for anisotropic nanoelectronic and optoelectronic applications.
	
	%%%%%%%%%%%%%%%%%%%%%%%%%%%%%%%%%%%%%%%%%%%%%%%%%%%%%%%%%%%%%%%%%%%%%%%%%%%%%
	\section{Methodology}
	
	First-principles calculations were performed using the SIESTA code \cite{Soler2002, Garcia2020}, which is based on the DFT formalism \cite{Kohn1965, Sanchez1997}. The exchange-correlation effects were described using the generalized gradient approximation (GGA) with the Perdew-Burke-Ernzerhof (PBE) functional \cite{Perdew1996}. Given that GGA/PBE systematically underestimates the electronic band gap, additional single-point calculations were carried out using the hybrid Heyd-Scuseria-Ernzerhof (HSE06) functional \cite{Heyd2003} as implemented in the HONPAS package \cite{Qin2015, Shang2020}. The electron-ion interactions were described by Troullier-Martins norm-conserving pseudopotentials \cite{Troullier1991} in the Kleinman-Bylander factored form \cite{Kleinman1982}, with valence electron configurations of 3\textit{d}$^{10}$4\textit{s}$^{2}$4\textit{p}$^{1}$ and 2\textit{s}$^{2}$2\textit{p}$^{3}$ for Ga and N atoms, respectively. An energy cutoff of 800~Ry and a double-$\zeta$ polarized (DZP) basis set composed of numerical atomic orbitals with finite range were employed in all calculations. The Brillouin zone was sampled using a $40 \times 40 \times 1$ Monkhorst-Pack $k$-point mesh for the unit cell and a $10 \times 10 \times 1$ mesh for the supercell calculations \cite{Monkhorst1976}. In the structural optimization, both the lattice vectors and atomic positions were fully relaxed until the maximum force acting on each atom was less than $10^{-3}$~eV/\AA\ and the difference in total energy was less than $10^{-5}$~eV. A vacuum spacing of 30~\AA\ was adopted along the direction perpendicular to the monolayer to prevent spurious interactions between periodic images.
	
	Phonon dispersion calculations were performed to assess the dynamical stability of the T-GaN monolayer. A $3 \times 3 \times 1$ supercell was employed with an energy convergence criterion of $10^{-5}$~eV, a force convergence criterion of 0.001~eV/\AA, and a mesh cutoff of 800~Ry. The acoustic sum rule was applied to the vibrational frequencies at the $\Gamma$ point. Additionally, AIMD simulations were performed using a $3 \times 3 \times 1$ supercell containing 36 atoms to assess the thermal stability of T-GaN at room temperature. The simulations were performed within the NVT ensemble using a time step of 1.0~fs and a Nos\'{e}-Hoover thermostat for temperature regulation. The system was evolved for 10~ps after heating to the target temperature \cite{tromer2020, montes2016, silva2021}.
	
	The mechanical properties were evaluated by applying uniaxial and biaxial strains to a $3 \times 3 \times 1$ supercell. For uniaxial strain, the lattice constants $a$ and $b$ were varied independently in increments of 0.5\% from their optimized values up to the end of the plastic region. The in-plane stiffness was obtained from \cite{Ozcelik2013, Topsakal2010}
	\begin{equation}
		C=\frac{1}{A} \times\left.\left(\frac{\partial^2 E_S}{\partial \varepsilon^2}\right)\right|_{\varepsilon\;=\;0} ,
	\end{equation}
	\noindent where $A$ is the area of the optimized supercell, $\varepsilon = \Delta l/l_0$ is the uniaxial strain with $l_0$ being the equilibrium supercell length and $\Delta l = l - l_0$, and $E_S$ is the strain energy calculated at each point by subtracting the equilibrium total energy from the total energy under strain. The in-plane bulk modulus was calculated as \cite{Majidi2017, Asadpour2015}
	\begin{equation}
		B=A \times\left.\left(\frac{\partial^2 E_S}{\partial A^2}\right)\right|_{A_{0}} ,
	\end{equation}
	\noindent where $A_0$ and $A$ represent the unit cell area at equilibrium and after the application of strain, respectively. The Poisson's ratio was obtained from the ratio of transverse strain ($\varepsilon_{\text{trans}}$) to axial strain ($\varepsilon_{\text{axial}}$) for small strain values \cite{Kang2011}
	\begin{equation}
		\nu=-\frac{\varepsilon_{\text{trans}}}{\varepsilon_{\text{axial}}} .
	\end{equation}
	
	The optical properties were computed by applying a standard external electric field of 1.0~V/\AA\ along the $x$-, $y$-, and $z$-directions separately. Using the Kramers-Kronig relation \cite{kramers1927, kronig1926} and Fermi's golden rule \cite{tignon1995}, the real ($\epsilon_1$) and imaginary ($\epsilon_2$) parts of the dielectric function were derived. The real part is given by
	\begin{equation}
		\epsilon_1(\omega)=1+\frac{1}{\pi}P\int_{0}^{\infty}d\omega'\frac{\omega'\epsilon_2(\omega')}{\omega'^{2}-\omega^{2}} \ \ ,
	\end{equation}
	\noindent where $P$ denotes the principal value of the integral over $\omega'$. The imaginary part, which accounts for interband optical transitions between the valence band (VB) and the conduction band (CB), is expressed as
	\begin{equation}
		\label{eq:epsilon_2}
		\epsilon_2(\omega)=\frac{4\pi^2}{\Omega\omega^2} \sum_{\substack{i\in \text{VB}\\ j\in \text{CB}}} \sum_{k}W_k \left | \rho_{ij} \right |^2 \delta (\epsilon_{kj}-\epsilon_{ki}-\hbar\omega) \ \ .
	\end{equation}
	\noindent In this equation, $\omega$ represents the photon frequency, $\left | \rho_{ij} \right |$ is the dipole transition matrix element, $W_k$ is the weight of the respective $k$-point in reciprocal space, and $\Omega$ is the system volume. From $\epsilon_1$ and $\epsilon_2$, additional optical quantities were obtained, including the absorption coefficient ($\alpha$), reflectivity ($R$), and refractive index ($\eta$)
	\begin{equation}
		\alpha(\omega)=\sqrt{2}\omega\left [  (\epsilon_1^{2}(\omega)+\epsilon_2^{2}(\omega))^{1/2}-\epsilon_1(\omega)\right ]^{1/2} ,
	\end{equation}
	\begin{equation}
		R(\omega)=\left [ \frac{(\epsilon_1(\omega)+i\epsilon_2(\omega))^{1/2}-1}{(\epsilon_1(\omega)+i\epsilon_2(\omega))^{1/2}+1} \right ]^2 ,
	\end{equation}
	and
	\begin{equation}
		\quad \eta(\omega)=\frac{1}{\sqrt{2}}\left [ (\epsilon_1^{2}(\omega)+\epsilon_2^{2}(\omega))^{1/2}+\epsilon_1(\omega) \right ]^{1/2} .
	\end{equation}
	
	%%%%%%%%%%%%%%%%%%%%%%%%%%%%%%%%%%%%%%%%%%%%%%%%%%%%%%%%%%%%%%%%%%%%%%%%%%%%%
	\section{Results and Discussion}
	
	\subsection{Structural Properties and Stability}
	
	\begin{figure}[t!]
		\centering
		\includegraphics[width=\linewidth]{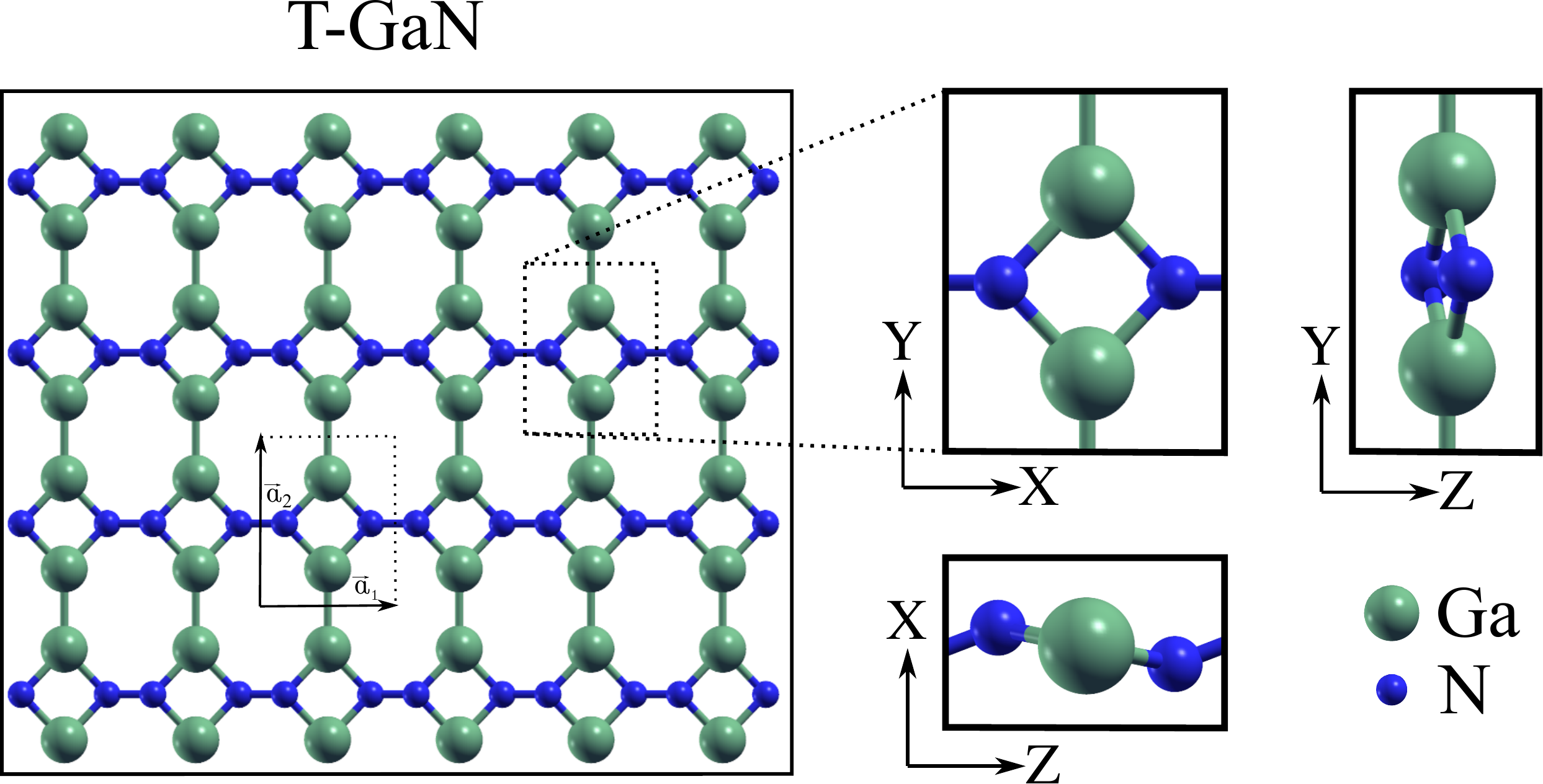}
		\caption{Atomic structure of the T-GaN monolayer. The main panel shows the top view (XY plane) with the unit cell vectors $\vec{a}_1$ and $\vec{a}_2$ indicated. The insets display magnified views along the XY, YZ, and XZ planes, highlighting the buckled geometry. Green and blue spheres represent Ga and N atoms, respectively.}
		\label{fig01}
	\end{figure}
	
	\begin{figure}[b!]
		\centering
		\includegraphics[width=0.7\linewidth]{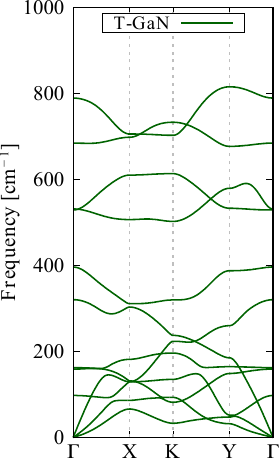}
		\caption{Phonon dispersion curves of the T-GaN monolayer calculated along the high-symmetry path $\Gamma$-X-K-Y-$\Gamma$ in the first Brillouin zone.}
		\label{fig02}
	\end{figure}
	
	The T-GaN monolayer adopts a tetragonal lattice composed of alternating four-membered and eight-membered rings, as illustrated in Figure~\ref{fig01}. The optimized structure exhibits a buckled geometry, in which the Ga and N atoms are displaced out of the basal plane, resulting in a non-planar configuration that accommodates mixed \textit{sp}$^{2}$/\textit{sp}$^{3}$ hybridization character. The unit cell contains four atoms (two Ga and two N) and is characterized by lattice parameters $a_1$ = 4.06~\AA\ and $a_2$ = 5.27~\AA. A distinctive feature of this T-GaN variant is the coexistence of three types of chemical bonds within the tetragonal motif. The optimized bond lengths are 1.95~\AA\ for the heteropolar Ga-N bonds, 1.55~\AA\ for the homoelemental N-N bonds, and 2.46~\AA\ for the homoelemental Ga-Ga bonds. In contrast, the all-Ga-N tetragonal variant \cite{Zhang2017} features exclusively heteropolar bonds with a reported length of approximately 1.83~\AA, resulting in a more symmetric configuration. The presence of Ga-Ga and N-N bonds in the present structure introduces additional bonding diversity, with the shorter N-N bonds reflecting stronger covalent interactions between nitrogen atoms, whereas the longer Ga-Ga bonds are consistent with the larger atomic radius and greater metallic character of gallium. This mixed bonding environment, which is absent in the all-Ga-N variant, is a direct consequence of the particular atomic arrangement within the 4-8 ring topology and is responsible for the anisotropic physical properties discussed in the following sections.
	
	\begin{figure*}[bp!]
		\centering
		\includegraphics[width=0.75\linewidth]{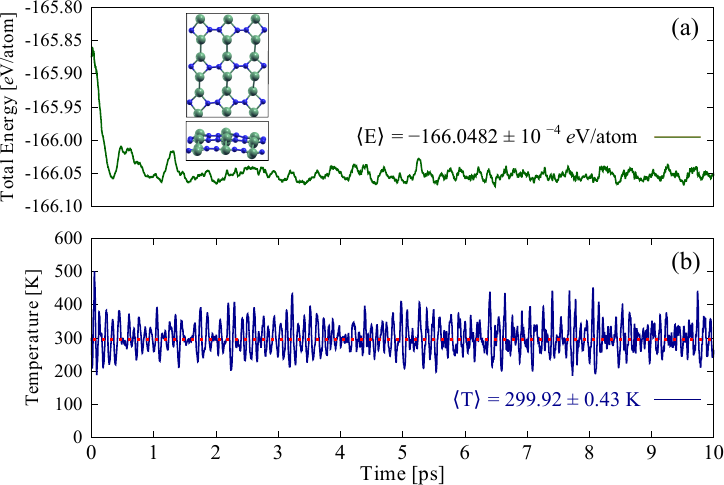}
		\caption{AIMD simulation results for the T-GaN monolayer at 300~K over 10~ps. (a) Time evolution of the total energy per atom. The insets show the structural snapshots at the end of the simulation (top and side views). (b) Time evolution of the temperature. The dashed red line indicates the target temperature of 300~K.}
		\label{fig03}
	\end{figure*}
	
	To assess the dynamical stability of T-GaN, we have calculated the phonon dispersion curves along the high-symmetry path $\Gamma$-X-K-Y-$\Gamma$ in the first Brillouin zone, as presented in Figure~\ref{fig02}. The unit cell containing four atoms gives rise to twelve phonon branches, of which three are acoustic, and nine are optical. We observe that all phonon frequencies are positive throughout the entire Brillouin zone, confirming the dynamical stability of the T-GaN monolayer. The three acoustic branches emanate from the $\Gamma$ point at zero frequency, as expected for a mechanically stable 2D structure. The highest optical frequency reaches approximately 815~cm$^{-1}$, which is consistent with the relatively strong Ga-N covalent bonds. Notably, a clear frequency gap is observed between the acoustic and optical branches, which is a characteristic feature of binary compounds with significant mass asymmetry between the constituent atoms \cite{Jain2014}.
	
	The thermal stability was further evaluated via AIMD simulations at 300~K for 10~ps. As shown in Figure~\ref{fig03}(a), the total energy per atom converges rapidly after an initial equilibration period and fluctuates around an average value of $-166.0482 \pm 10^{-4}$~eV/atom, indicating robust energetic stability at room temperature. Figure~\ref{fig03}(b) presents the temperature evolution, which oscillates around $299.92 \pm 0.43$~K, confirming the effectiveness of the Nos\'{e}-Hoover thermostat. The insets in Figure~\ref{fig03}(a) display snapshots of the structure at the end of the simulation, revealing that no significant structural distortions or bond-breaking events occurred during the molecular dynamics trajectory. Together, the phonon and AIMD results provide strong evidence that the T-GaN monolayer is both dynamically and thermally stable under ambient conditions. This finding is particularly noteworthy given that the present structure contains homoelemental Ga-Ga and N-N bonds, which are absent in the conventional hexagonal and in the all-Ga-N tetragonal phases \cite{Zhang2017, Yong2017}. The confirmed stability of a GaN monolayer with such mixed bonding character suggests that the design space for stable 2D group III-V materials may be broader than previously assumed, extending beyond configurations with exclusively heteropolar bonding.

	\subsection{Mechanical Properties}
	
	\begin{figure*}[b!]
		\centering
		\includegraphics[width=0.75\linewidth]{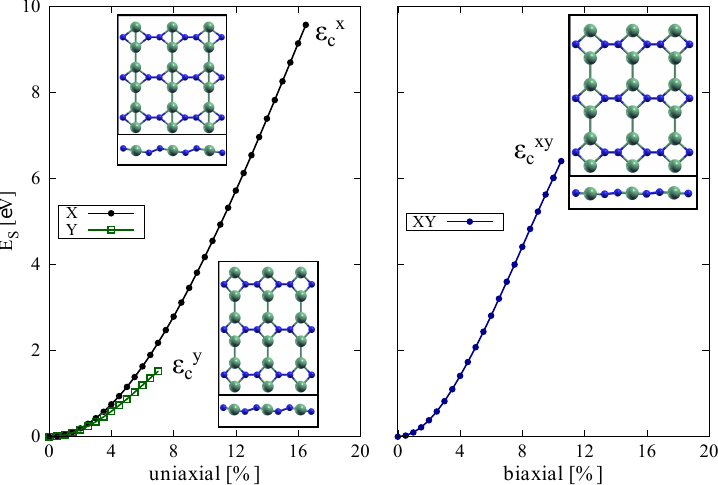}
		\caption{Strain energy ($E_S$) as a function of applied strain for the T-GaN monolayer. The left panel shows the uniaxial strain along the $x$- (black) and $y$-directions (green), and the right panel shows the biaxial strain along the $xy$-direction (blue). The critical strains $\varepsilon_c^x$, $\varepsilon_c^y$, and $\varepsilon_c^{xy}$ are indicated. The insets show structural snapshots at the corresponding critical strain values.}
		\label{fig08}
	\end{figure*}
	
	The mechanical response of T-GaN was characterized by applying uniaxial and biaxial strains, and the resulting strain energy curves are presented in Figure~\ref{fig08}. The left panel of Figure~\ref{fig08} shows the strain energy as a function of uniaxial strain applied along the $x$- and $y$-directions separately. We observe a pronounced mechanical anisotropy, with the $x$-direction exhibiting a considerably stiffer response than the $y$-direction. The critical strain, defined as the point beyond which the strain energy deviates from quadratic behavior, indicating the onset of plastic deformation, was estimated at approximately $\varepsilon_c^x$ = 16.5\% for the $x$-direction and $\varepsilon_c^y$ = 7.0\% for the $y$-direction. This substantial difference reflects the distinct bonding environments along the two crystallographic axes in the tetragonal lattice. Along the $x$-direction, the applied strain is primarily accommodated by the homoelemental N-N bonds, which possess a strong covalent character as evidenced by the high electron localization in the bonding region (see Section~3.3), thus conferring greater resistance to deformation. In contrast, along the $y$-direction, the strain acts predominantly on the homoelemental Ga-Ga bonds, which exhibit a more delocalized, metallic-like character and lower bond strength, leading to an earlier onset of plastic deformation. The right panel of Figure~\ref{fig08} presents the biaxial strain energy curve, which yields a critical strain of approximately $\varepsilon_c^{xy}$ = 10.5\%.
	
	From the quadratic region of the strain energy curves, the in-plane stiffness values were extracted as $C_x$ = 66.2~N/m and $C_y$ = 49.1~N/m for the $x$- and $y$-directions, respectively, and the biaxial in-plane bulk modulus was obtained as $B$ = 43.6~N/m. The Poisson's ratio was calculated as $\nu_{xy}$ = 0.12 and $\nu_{yx}$ = 0.15. These values indicate that T-GaN is mechanically softer and more anisotropic than its hexagonal counterpart g-GaN, for which the in-plane stiffness has been reported as approximately 110~N/m \cite{Peng2013}. The enhanced anisotropy in T-GaN originates directly from the reduced symmetry of the tetragonal lattice, which introduces direction-dependent bonding configurations absent in the hexagonal phase. The insets in Figure~\ref{fig08} display structural snapshots at the critical strain values, illustrating the distinct deformation mechanisms along each direction.
	
	\subsection{Electronic Properties}
	
\begin{figure}[t!]
		\centering
		\includegraphics[width=1.0\linewidth]{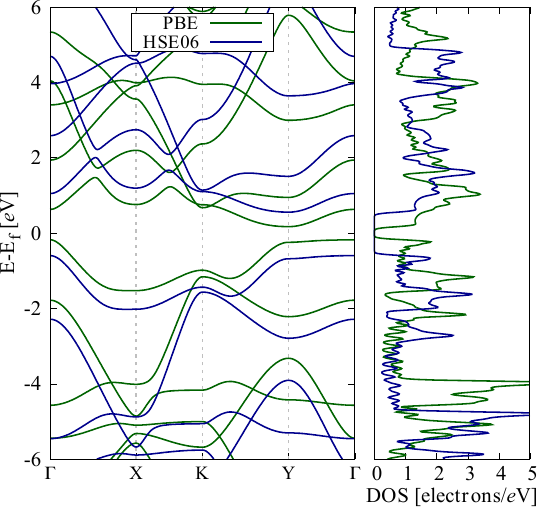}
		\caption{Electronic band structure of T-GaN calculated along the high-symmetry path $\Gamma$-X-K-Y-$\Gamma$ at the PBE (green) and HSE06 (blue) levels of theory. The right panel shows the corresponding density of states (DOS). The Fermi level is set to zero.}
		\label{fig04}
	\end{figure}

	The electronic band structure of T-GaN was calculated along the high-symmetry path $\Gamma$-X-K-Y-$\Gamma$ at both the PBE and HSE06 levels of theory, as presented in Figure~\ref{fig04} alongside the corresponding density of states (DOS). We observe that T-GaN is a semiconductor with an indirect band gap, where the valence band maximum (VBM) is located near the Y point, and the conduction band minimum (CBM) is located at the $\Gamma$ point. The calculated band gap is 0.35~eV at the PBE level and 1.15~eV with the HSE06 hybrid functional. These values are considerably lower than the PBE band gaps reported for other tetragonal GaN variants, namely approximately 1.89~eV for the all-Ga-N configuration \cite{Zhang2017} and approximately 1.6~eV for the GaN Haeckelite monolayer \cite{CamachoMojica2015}. This reduction may be attributed to the presence of Ga-Ga and N-N bonds in the present structure, which introduce additional states near the band edges and modify the orbital hybridization pattern compared to configurations with exclusively heteropolar bonding. As expected, the HSE06 functional yields a wider band gap due to the partial incorporation of exact exchange, which partially corrects the well-known self-interaction error inherent to semilocal functionals. The overall qualitative features of the band structure, including band curvatures and dispersion character, remain consistent across both levels of theory, indicating that the PBE description captures the essential physics of the electronic structure, while the HSE06 correction provides a more accurate quantitative estimate of the fundamental gap.
	
	\begin{figure*}[t!]
		\centering
		\includegraphics[width=0.7\linewidth]{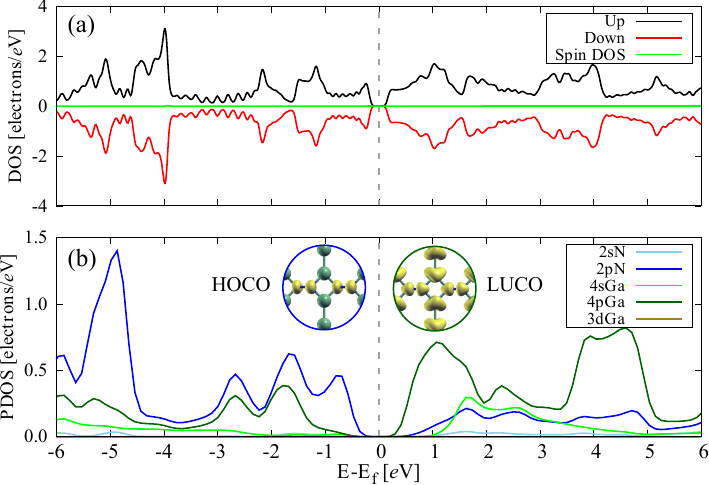}
		\caption{(a) Spin-polarized density of states showing the spin-up (black), spin-down (red), and net spin DOS (green) of T-GaN. (b) Projected density of states (PDOS) decomposed into orbital contributions from N (2\textit{s}, 2\textit{p}) and Ga (4\textit{s}, 4\textit{p}, 3\textit{d}) atoms. The insets display the spatial distribution of the HOCO and LUCO orbitals. The Fermi level is indicated by the dashed gray line.}
		\label{fig05}
	\end{figure*}
	
	To gain further insight into the electronic character of T-GaN, we have calculated the spin-polarized density of states and the projected density of states (PDOS), as shown in Figure~\ref{fig05}. The spin-resolved DOS in Figure~\ref{fig05}(a) reveals a perfectly symmetric distribution between spin-up and spin-down channels, with a vanishing net spin density across the entire energy range. This result confirms the nonmagnetic ground state of T-GaN, which is consistent with the absence of unpaired electrons in the fully bonded tetragonal configuration. The PDOS analysis presented in Figure~\ref{fig05}(b) provides a detailed orbital decomposition of the electronic states. The valence band is predominantly composed of nitrogen 2\textit{p} orbitals, with a smaller contribution from gallium 4\textit{p} states, while the conduction band is governed primarily by gallium 4\textit{s} and 4\textit{p} orbitals. This orbital composition indicates a significant charge transfer from Ga to N atoms, consistent with the ionic character of the Ga-N bond. The insets in Figure~\ref{fig05}(b) display the spatial distribution of the highest occupied crystalline orbital (HOCO) and the lowest unoccupied crystalline orbital (LUCO), which are localized on the nitrogen and gallium sublattices, respectively.
	
	\begin{figure}[b!]
		\centering
		\includegraphics[width=0.8\linewidth]{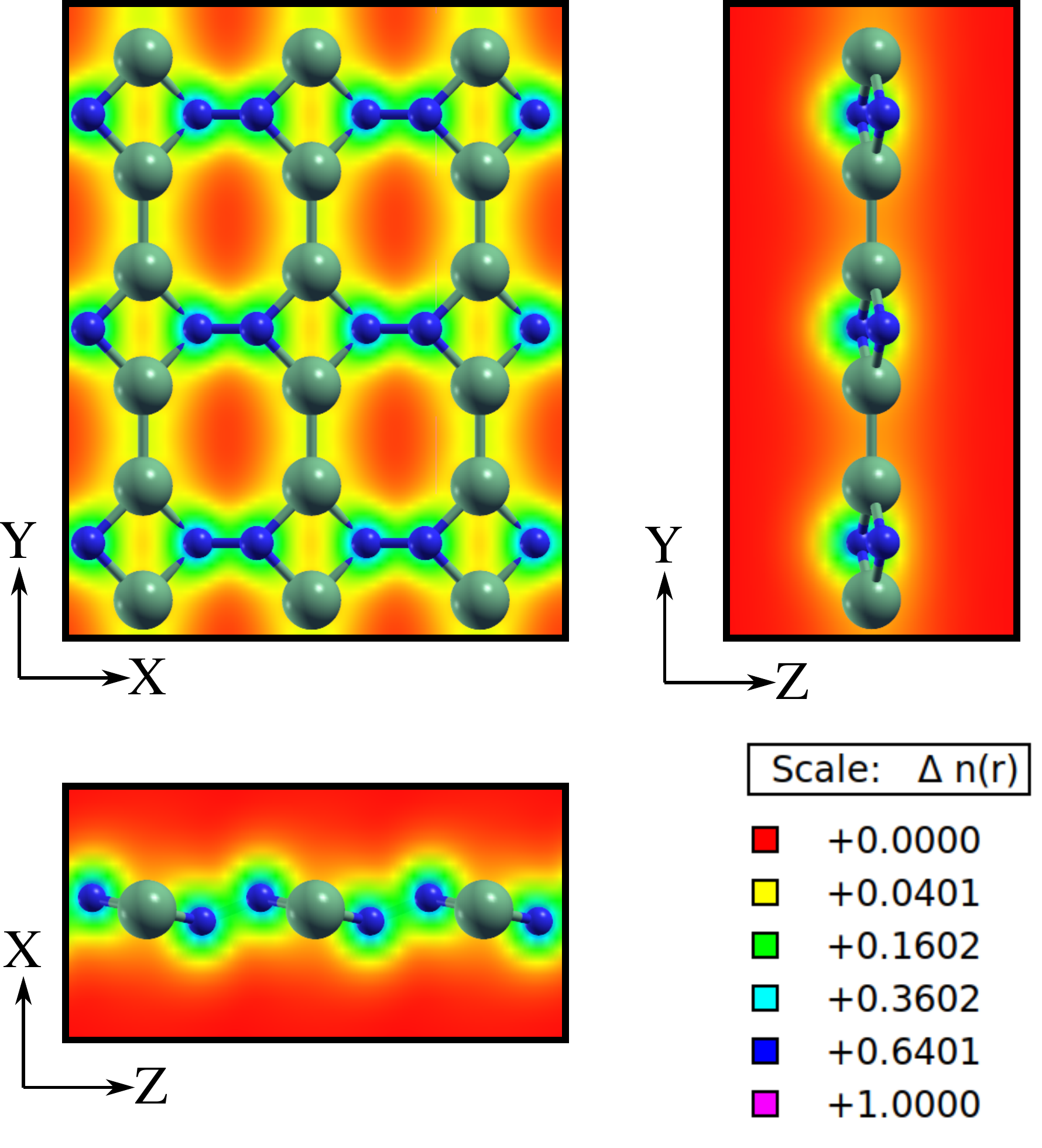}
		\caption{Electron localization function (ELF) of the T-GaN monolayer plotted along the XY (top left), YZ (top right), and XZ (bottom) crystallographic planes. The color scale represents electron localization, ranging from 0.0 (fully delocalized, red) to 1.0 (fully localized, purple).}
		\label{fig06}
	\end{figure}
	
	\begin{figure*}[b!]
		\centering
		\includegraphics[width=0.75\linewidth]{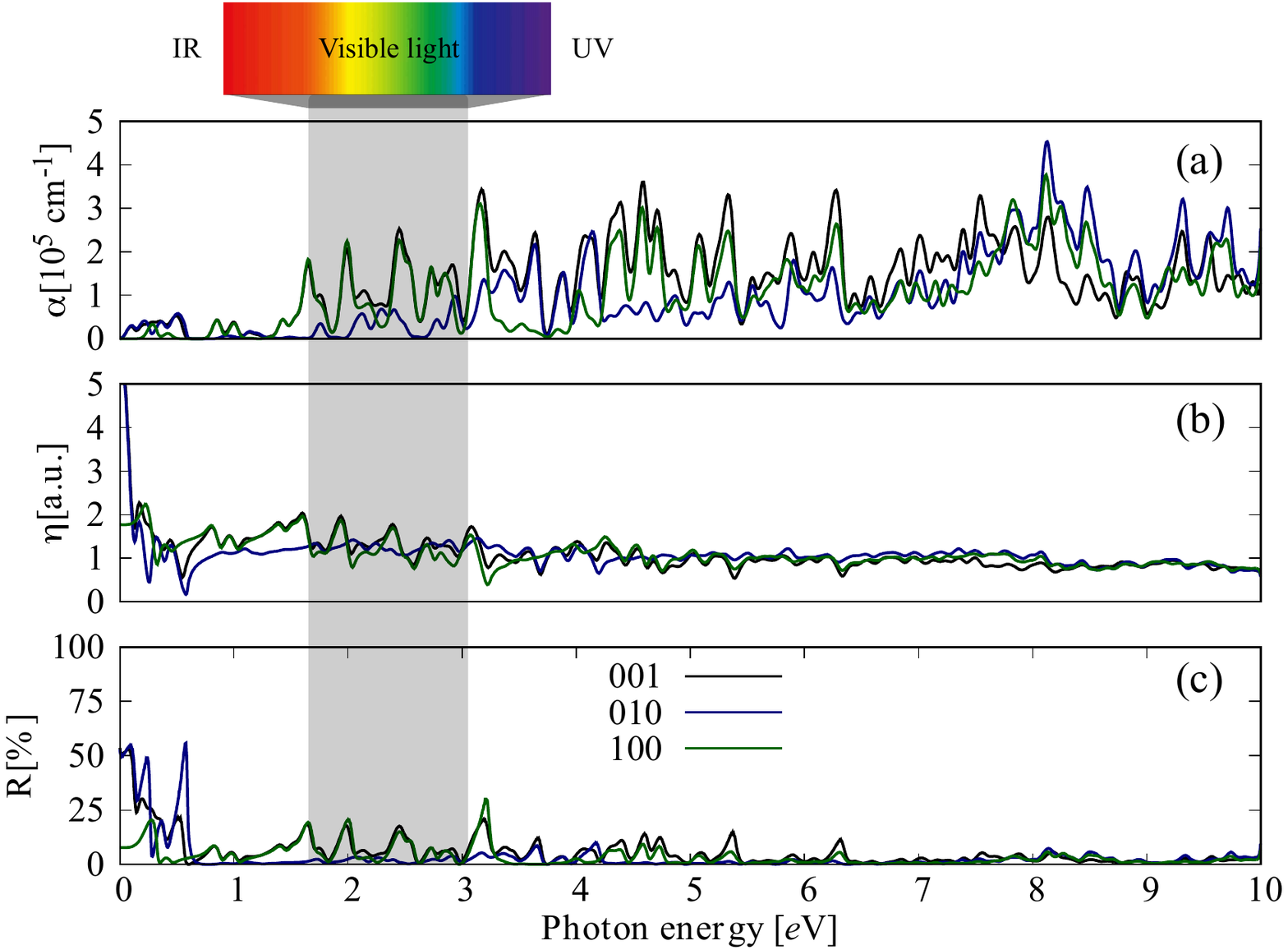}
		\caption{Optical properties of the T-GaN monolayer calculated along the 001 (black), 010 (blue), and 100 (green) directions. (a) Absorption coefficient $\alpha$. (b) Refractive index $\eta$. (c) Reflectivity $R$. The shaded region indicates the visible light range.}
		\label{fig07}
	\end{figure*}
	
	The electron localization function (ELF) was computed to characterize the nature of the chemical bonding in T-GaN, as illustrated in Figure~\ref{fig06}. The ELF maps are presented along three crystallographic planes (XY, YZ, and XZ) and employ a color scale ranging from 0 (red, fully delocalized) to 1 (purple, fully localized). We observe that the highest electron localization (cyan and blue regions) is concentrated around the nitrogen atoms and along the Ga-N bonds, while the gallium sites exhibit low electron density (red and orange regions). This distribution is characteristic of a polar covalent bond with a significant ionic contribution, in which the more electronegative nitrogen atoms attract electron density from neighboring gallium atoms. Additionally, the ELF maps reveal distinct bonding signatures for the three types of bonds present in the structure. The homoelemental N-N bonds exhibit a relatively high electron localization in the bonding region, consistent with a strong covalent interaction between nitrogen atoms. In contrast, the homoelemental Ga-Ga bonds display lower electron localization, reflecting the more delocalized character of the gallium-gallium interaction. The Ga-N bonds exhibit an asymmetric electron distribution shifted toward nitrogen, consistent with the electronegativity difference between the two elements. These contrasting bonding signatures provide a direct electronic-structure basis for the pronounced mechanical anisotropy discussed in Section~3.2, where the stiffer $x$-direction is governed by the strongly covalent N-N bonds and the softer $y$-direction is controlled by the weaker Ga-Ga bonds. The ELF pattern is also consistent with the PDOS analysis and further corroborates the charge-transfer mechanism that governs the electronic structure of T-GaN.
	
	\subsection{Optical Properties}
	
	The optical properties of T-GaN were evaluated along the three crystallographic directions (001, 010, and 100), as presented in Figure~\ref{fig07}. The absorption coefficient $\alpha(\omega)$ is shown in Figure~\ref{fig07}(a), where the shaded region indicates the visible light range (approximately 1.63 to 3.26~eV). We observe that the absorption onset occurs at relatively low photon energies, consistent with the material's calculated band gap. The 001 direction (perpendicular to the monolayer plane) exhibits a markedly different absorption profile compared to the in-plane directions (100 and 010), with notable absorption activity extending into the infrared and visible regions. The in-plane directions exhibit similar absorption characteristics, with significant absorption beginning in the visible range and intensifying in the ultraviolet. This directional anisotropy in the optical absorption reflects the anisotropic electronic structure and the distinct dipole transition matrix elements associated with in-plane and out-of-plane polarization.
	
	The refractive index $\eta(\omega)$ is presented in Figure~\ref{fig07}(b). The static refractive index $\eta(0)$ exhibits anisotropy between the three directions, with the 001 direction yielding values of approximately 2.0--2.5, while the 100 and 010 directions present lower values of approximately 1.5--1.8. These static values are of practical relevance for optoelectronic device design, as they determine the light propagation characteristics within the material.
	
	The reflectivity $R(\omega)$ is displayed in Figure~\ref{fig07}(c). The 001 direction presents a prominent reflectivity peak of approximately 50\% at low photon energies, while the in-plane reflectivity remains below 25\% across most of the spectrum. The overall reflectivity decreases with increasing photon energy in all directions, reaching values below 5\% in the deep ultraviolet region. These results suggest that T-GaN may be considered a transparent material in the ultraviolet regime while exhibiting moderate reflectivity in the visible and infrared ranges along specific directions.
	
	%%%%%%%%%%%%%%%%%%%%%%%%%%%%%%%%%%%%%%%%%%%%%%%%%%%%%%%%%%%%%%%%%%%%%%%%%%%%%
	\section{Conclusions}
	
	In summary, we have performed a comprehensive first-principles investigation of the structural, mechanical, electronic, and optical properties of the T-GaN monolayer using DFT within the GGA/PBE and HSE06 formalisms. The dynamical and thermal stability of T-GaN was confirmed by phonon dispersion calculations, which showed the absence of imaginary frequencies, and by AIMD simulations at 300~K, which demonstrated robust structural integrity over a 10~ps trajectory.
	
	Notably, the mechanical characterization revealed a pronounced in-plane anisotropy, with critical strains differing by approximately a factor of 2.4 between the $x$- and $y$-directions, which originates from the distinct bonding environments within the four- and eight-membered ring motifs. The electronic structure analysis revealed that T-GaN is a nonmagnetic semiconductor with an indirect band gap of 0.35~eV (PBE) and 1.15~eV (HSE06), where the valence band is dominated by nitrogen 2\textit{p} states and the conduction band by gallium 4\textit{s} and 4\textit{p} orbitals. The ELF analysis confirmed the polar covalent nature of the Ga-N bonding with significant charge transfer from gallium to nitrogen, and revealed distinct bonding signatures for the three types of bonds that characterize this structural variant. The optical properties exhibited considerable directional anisotropy, with the out-of-plane response differing markedly from the in-plane response across the absorption coefficient, refractive index, and reflectivity.
	
	These findings demonstrate that the tetragonal topology with mixed bonding character endows T-GaN with highly anisotropic physical properties that are absent in the hexagonal phase and distinct from those of the all-Ga-N tetragonal variant, suggesting that T-GaN may be a promising candidate for applications requiring directional selectivity in mechanical response, electronic transport, or optical absorption. Moreover, the demonstrated dynamical and thermal stability of a GaN monolayer containing homoelemental Ga-Ga and N-N bonds is a significant result in itself, as it indicates that stable 2D group III-V structures are not restricted to configurations with exclusively heteropolar bonding. This observation opens a promising avenue for the systematic exploration of alternative atomic arrangements within non-hexagonal topologies, potentially expanding the catalog of viable 2D III-V materials with tailored properties. The comprehensive characterization reported in this study contributes to the growing understanding of non-hexagonal 2D group III-V materials and may guide future experimental efforts toward the realization of T-GaN-based nanodevices.
	
	%%%%%%%%%%%%%%%%%%%%%%%%%%%%%%%%%%%%%%%%%%%%%%%%%%%%%%%%%%%%%%%%%%%%%%%%%%%%%
	\section*{Acknowledgments}
	D.S.G. acknowledges financial support from the Coordena\c{c}\~{a}o de Aperfei\c{c}oamento de Pessoal de N\'{i}vel Superior (CAPES). M.L.P.J. acknowledges financial support from FAPDF (grant 00193-00001807/2023-16), CNPq (grants 444921/2024-9 and 308222/2025-3), and CAPES (grant 88887.005164/2024-00). S.A acknowledges financial support from CNPq ( grants 303069/2021-0) and INCT - Nanomateriais de Carbono .
	
	\bibliography{refs}
	
\end{document}